%% file: main.tex
\documentclass[10pt,conference]{IEEEtran}
\IEEEoverridecommandlockouts

\usepackage{todonotes}
\usepackage{cite}
\usepackage{amsmath,amssymb,amsfonts}
\usepackage{algorithmic}
\usepackage{graphicx}
\usepackage{textcomp}
\usepackage{xcolor}
\usepackage{multirow}
\usepackage{booktabs}
\usepackage{array}

\usepackage{hyperref}
\hypersetup{
    colorlinks=true,
    allcolors=blue,
    pdftitle={Dynamic Risk Assessment by Bayesian Attack Graphs and Process Mining},
    pdfpagemode=FullScreen,
}

\newcommand{\revision}{\textcolor{black}}

\def\BibTeX{{\rm B\kern-.05em{\sc i\kern-.025em b}\kern-.08em
    T\kern-.1667em\lower.7ex\hbox{E}\kern-.125emX}}

\usepackage{pbalance}

\makeatletter
\newcommand{\linebreakand}{%
  \end{@IEEEauthorhalign}
  \hfill\mbox{}\par
  \mbox{}\hfill\begin{@IEEEauthorhalign}
}
\makeatother
\newcolumntype{P}[1]{>{\centering\arraybackslash}p{#1}}

\begin{document}
\title{Dynamic Risk Assessment by \\Bayesian Attack Graphs and Process Mining
\thanks{This work was supported by Agenzia per la cybersicurezza nazionale under the programme for promotion of XL cycle PhD research in cybersecurity – C83C24000790001. The views expressed are those of the authors and do not represent the funding institution.}
}

\author{%
\IEEEauthorblockN{Francesco Vitale}
\IEEEauthorblockA{\textit{DIETI} \\
\textit{University of Naples Federico II}\\
Naples, Italy \\
francesco.vitale@unina.it}
\and
\IEEEauthorblockN{Simone Guarino}
\IEEEauthorblockA{\textit{Department of Engineering} \\
\textit{University Campus Bio-Medico of Rome}\\
Rome, Italy \\
s.guarino@unicampus.it}
\and
\IEEEauthorblockN{Stefano Perone}
\IEEEauthorblockA{\textit{Department of Engineering} \\
\textit{University Campus Bio-Medico of Rome}\\
Rome, Italy \\
s.perone@unicampus.it}
\linebreakand
\IEEEauthorblockN{Massimiliano Rak}
\IEEEauthorblockA{\textit{DIETI} \\
\textit{University of Naples Federico II}\\
Naples, Italy \\
massimiliano.rak@unina.it}
\and
\IEEEauthorblockN{Nicola Mazzocca}
\IEEEauthorblockA{\textit{DIETI} \\
\textit{University of Naples Federico II}\\
Naples, Italy \\
nicola.mazzocca@unina.it}
}

\maketitle

\begin{abstract}
While attack graphs are useful for identifying major cybersecurity threats affecting a system, they do not provide operational support for determining the likelihood of having a known vulnerability exploited, or that critical system nodes are likely to be compromised. In this paper, we perform dynamic risk assessment by combining Bayesian Attack Graphs (BAGs) and online monitoring of system behavior through process mining. Specifically, the proposed approach applies process mining techniques to characterize malicious network traffic and derive evidence regarding the probability of having a vulnerability actively exploited. This evidence is then provided to a BAG, which updates its conditional probability tables accordingly, enabling dynamic assessment of vulnerability exploitation. We apply our method to a cybersecurity testbed instantiating several machines deployed on different subnets and affected by several CVE vulnerabilities. The testbed is stimulated with both benign traffic and malicious behavior, which simulates network attack patterns aimed at exploiting the CVE vulnerabilities. The results indicate that our proposal effectively detects whether vulnerabilities are being actively exploited, allowing for an updated assessment of the probability of system compromise.

\end{abstract}


\input{sections/introduction}
\input{sections/related_work}
\input{sections/approach}
\input{sections/evaluation}
\input{sections/conclusion}


\bibliographystyle{IEEEtran}
\bibliography{bibliography}

\end{document}

%% file: sections/introduction.tex
\section{Introduction}
\label{sec:intro}
The complexity of modern cyber-physical systems exposes them to several vulnerabilities that may be exploited by malicious users to impair their services and/or intrude on their network. To evaluate such risks, various assessment methods have been proposed based on graph theory and quantitative measures extracted from vulnerability databases, such as the Common Vulnerabilities and Exposures (CVE) database~\cite{arat2026graphbasedriskassiot}.

A widespread approach for risk assessment involves the use of attack graphs, which are abstract descriptions of multi-step attacks aimed at compromising the different nodes of the network. By combining attack graphs and known vulnerabilities, a quantitative evaluation of the risk can be obtained~\cite{viticchie2025attackgraphframework}. An effective method to quantitatively evaluate the risk of system compromise is the use of Bayesian Attack Graphs (BAGs), which model risk likelihood through conditional probabilities~\cite{perone2025vulnasscvss}. Specifically, to each node of a BAG is assigned a Conditional Probability Table (CPT), which reports the probability that the node is compromised based on the compromise of parent nodes linked to the attack path that can be followed.

While BAGs can dynamically update posterior probabilities based on isolated alerts, their underlying CPTs traditionally rely on static metrics, such as CVSS scores. Furthermore, they lack a mechanism to continuously ingest complex behavioral evidence. In this paper, we aim to support dynamic risk assessment by extending BAGs with online diagnoses provided by process mining. \revision{This research area allows extracting process models from network event data that capture the packet-level sequencing of network protocols~\cite{vitale2025networktrafficanalysisprocess}. Hence, process mining techniques enable 1) packet-level inspection and modeling of network protocols, and 2) evaluating the similarity of new behavior with the prior characterization. In view of this two-fold utility,} our proposed approach leverages process mining techniques to characterize malicious network traffic and derive evidence regarding the probability of having a vulnerability actively exploited. This evidence is then provided to a BAG, which updates its conditional probability tables accordingly, enabling dynamic assessment of vulnerability exploitation.

The novelties and contributions of our proposal are thus the following:
\begin{itemize}
    \item Process mining-based monitoring of vulnerability exploitation from the analysis of benign and malicious traffic.
    \item Dynamic risk assessment through BAGs enhanced with the diagnoses provided by process mining-based monitoring.
\end{itemize}

We test our method with a cybersecurity testbed instantiating a network of nodes with known CVE vulnerabilities. The testbed is vulnerable to a multi-step attack captured in a BAG. The experiments involved characterizing malicious traffic associated with CVE vulnerabilities exploitation and providing dynamic risk assessment supported by process mining-based diagnoses. 


%% file: sections/related_work.tex
\section{Background and Related Work}
\label{sec:related_work}

\subsection{Dynamic Risk Assesment}
BAGs provide a powerful framework for vulnerability assessment by modeling exploitation probabilities of vulnerabilities affecting network infrastructures. Formally, a BAG is defined by the tuple $BAG = (V, E, U, P)$, where nodes $V$ represent the attacker’s access privileges on specific devices, namely \emph{security conditions}, edges $E$ denote the existence of exploits that enable transitions between nodes, $U$ is the set of software vulnerabilities affecting each device, and $P$ represents the probability of successfully exploiting each vulnerability. In particular, each edge is defined by the tuple $e_i = (v_i, v_j, u, P(u)) \in E$, where $P(u)$ denotes the probability of exploiting vulnerability $u$ to transition from condition $v_i$ to $ v_j$. These probabilities populate the Conditional Probability Tables (CPT) at each node, which encode the causal dependencies among different security conditions.

In recent years, several studies have investigated methods for accurately estimating $P(u)$ given a set of vulnerabilities $U$ affecting a network infrastructure. Poolsappasit et al. \cite{5936075} proposed a dynamic risk management framework based on BAGs, where the exploitation probability of vulnerabilities is evaluated using the Common Vulnerability Scoring System (CVSS) Exploitability sub-score, which captures how easily a vulnerability can be exploited. Similarly, Guarino et al. \cite{guarino2024holistic} proposed a BAG-based vulnerability assessment approach that considers both the Exploitability and Temporal metrics of CVSS, thus providing a more comprehensive characterization of vulnerability severity by accounting for the availability of patches and known exploits.
Although CVSS is widely adopted, its scores have been demonstrated not being directly correlated to the exploitation probability \cite{10076946}. To overcome this limitation, recent research has focused on the Exploit Prediction Scoring System (EPSS), a metric that explicitly estimates the probability that a given vulnerability will be exploited in the wild within 30 days. EPSS scores are updated daily, enabling the dynamic evaluation of vulnerability exploitation probability over time. Cheimonidis et al. \cite{11130107} introduced a novel framework combining Bayesian Networks, Markov chains, and EPSS values to quantify the probabilistic transitions between compromised states and to measure the likelihood of reaching critical assets. Similarly, Yadav \cite{yadav2025extent} proposed an approach to estimate transition probabilities within attack graphs by integrating both CVSS and EPSS, thereby capturing both the severity of vulnerabilities and their time-varying likelihood of exploitation.

However, despite these advances, existing approaches still face challenges in updating posterior probabilities based on live evidence from ongoing cyber-attacks. Such attacks usually follow well-defined stages in the exploitation of known vulnerabilities, which can be traced and recognized through pattern-based detection systems. Therefore, there is a need to feed BAGs with evidence of active vulnerability exploitation to dynamically update the posterior probabilities of system compromise. To date, most existing works tend to inject synthetic evidence on BAG nodes, without explicitly modeling the concrete sequence of steps required by an attacker to exploit a given vulnerability 
\cite{sahu2023inferring, cerotti2025dynamic}.
\begin{figure*}[t]
    \centering
    \includegraphics[width=0.825\textwidth]{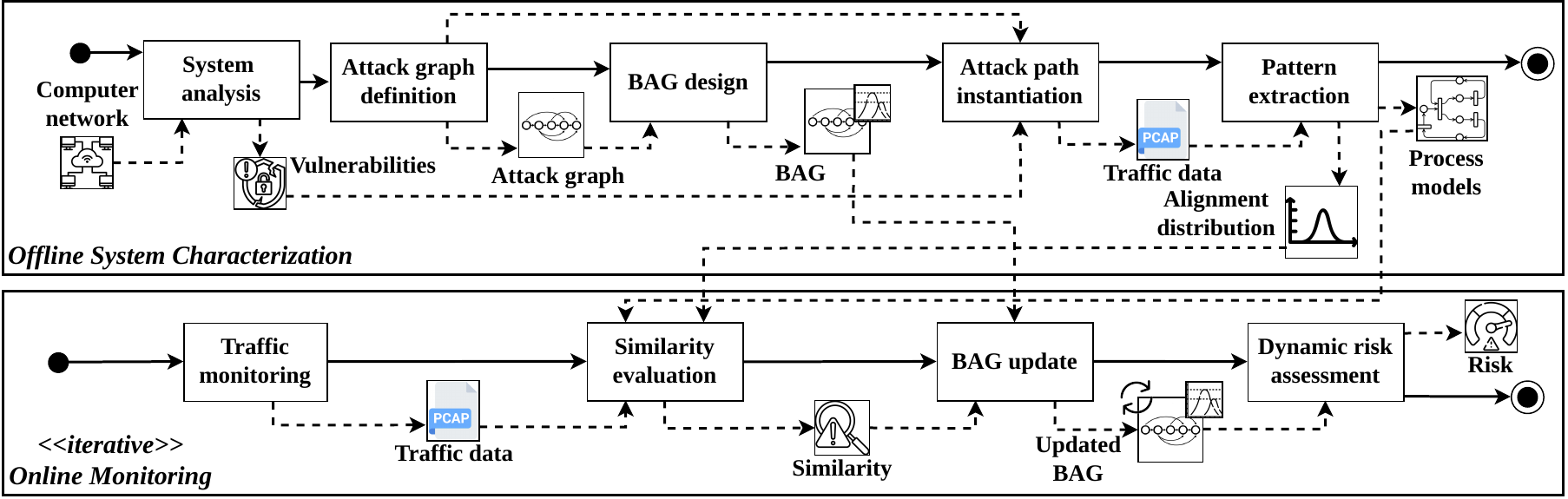}
    \caption{The proposed approach for dynamic risk assessment through process mining and BAGs.}
    \label{METHOD}
\end{figure*}

\subsection{Process Mining}
Process mining is able to bridge data science with process science by characterizing actual behavior from monitored events through process models and verifying the conformity of new events with that characterization~\cite{aalst2022pmhb}. In particular, process mining mainly involves two sets of algorithms: process discovery and conformance checking. Both types of algorithms deal with so-called event logs. An event log is a multiset of traces: $L\in\mathcal{B}(\mathcal{A}^*)$, where $\mathcal{A}$ indicates a set of events, $\mathcal{A}^*$ the set of all ordered sequences of events (i.e., traces), and $\mathcal{B}(\mathcal{A}^*)$ the universe of multisets that can be built with $\mathcal{A}^*$. On the one hand, a process discovery algorithm $\gamma$ can discover a process model $N$ from $L$, i.e., $\gamma(L)=N$. Given a trace $\sigma\in\mathcal{A}^*$, a conformance checking algorithm is able to obtain a set of diagnoses $d_\sigma\in\mathbb{R}^{|\mathcal{A}|+1}$ by comparing $\sigma$ with $N$. In particular, the diagnoses $d_\sigma$ contain local information about activities aligning with the description in $N$ and a fuzzy value that quantifies global conformance to the overall process model, i.e., the fitness~\cite{vitale2025cfadpm}.

Process mining has been applied to network traffic of several protocols to analyze message patterns and uncover anomalous behavior. Its advantages include providing packet-level inspection and explicit capture of control-flow aspects of network protocols, and verifying whether new traffic is aligned with historical patterns and potentially diagnose specific anomalies. For example, Bustos-Jimenez et al.~\cite{bustosjimenez2014applyingpmdns} utilized process discovery to analyze the DNS traffic of a large dataset comprising DNS queries to different Internet services. Their findings highlight the potential of process mining to diagnose the most frequent network activities under normal conditions and verify anomalies in presence of network intrusions, such as botnet attacks. Wakup and Desel~\cite{wakup2015analyzingtcpip} developed an automated tool to convert TCP traffic into a process model through pre-processing and event-log extraction from network data; their work showed that it was possible to reconstruct a legacy protocol from TCP traffic. Vitale et al.~\cite{vitale2025networktrafficanalysisprocess} developed a dedicated pipeline to streamline and segment the traffic of network protocols into different states to build more precise and interpretable process models.

\revision{The discussion above outlines that process mining is able to perform packet-level inspection of various network protocols and evaluate whether new network traffic complies with the extracted patterns. Combining process mining with} the ability of BAGs to dynamically assess the risk associated with network intrusions, our method allows for early diagnosis of known attack paths and dynamically assesses the risk of security breaches.

%% file: sections/approach.tex
\section{Proposed Approach}
\label{sec:approach}
In this section, we present the proposed approach, depicted in Fig. \ref{METHOD}. The approach is structured in two phases: offline system characterization and online monitoring. 

\subsection{Offline System Characterization}
The offline phase begins by characterizing the target computer network in terms of its devices, installed software, and associated cyber vulnerabilities. First, \textbf{system analysis} aims to discover all reachable devices in the network, perform a comprehensive enumeration of the applications installed on each node, and, finally, associate with every device the corresponding list of known vulnerabilities that affect it. These vulnerabilities are automatically identified using network scanning tools such as Nmap and Greenbone OpenVAS, which retrieve vulnerability data from the CVE database. 
Subsequently, \textbf{attack graph generation definition} leverages this network characterization and vulnerability inventory to model all viable Attack Paths (APs) from one or multiple entry points to target devices via vulnerability exploitation. The attack graph construction employs pre-condition and post-condition modeling for each vulnerability, defining exploitation prerequisites and resulting privilege escalation. The resulting model constitutes a directed graph where each node represents the security condition of a specific host (guest, user, or root access), and each edge represents an elementary attack action exploiting a single vulnerability. Subsequently, the \textbf{BAG design} extends the AG by assigning an exploitation probability to each vulnerability. These probability values are dynamically updated within the CPTs at each node during the Online Monitoring phase.

After generating the BAG, \textbf{attack path instantiation} involves capturing malicious behavioral patterns in response to the exploitation of the vulnerabilities resulting from network analysis. To this aim, known attacks are replicated against system nodes to exploit the corresponding vulnerability. The replication of the attack involves executing a series of actions that cause anomalous network traffic. After recording the traffic data generated during the execution of the attack, \textbf{pattern extraction} characterizes the specific packet-level network patterns through the approach proposed in~\cite{vitale2025networktrafficanalysisprocess}, which reports a structured method for extracting process models from traffic data. Let $\mathcal{W}=\{W_1,\dots,W_\alpha\}$ indicate the set of $\alpha$ vulnerabilities found from network analysis, and $\mathcal{T}=\{T_1,\dots,T_\alpha\}$ indicate the set of traffic data obtained by exploiting the $\alpha$ vulnerabilities. Without loss of generality, we will consider that each node of the computer network is associated with a single vulnerability. To characterize the pattern of the $i$-th vulnerability $W_i$, the method involves the application of three steps: 1) feature extraction to obtain numerical statistics from the raw network data, 2) state-space characterization to cluster network patterns into distinct groups, and 3) event log extraction to turn the pre-processed traffic data $T_i$ into $\beta$ event logs $\mathcal{L}_i=\{L_1,\dots,L_\beta:L_j\in\mathcal{B}(\mathcal{A}^*)\}$, where $\mathcal{A}$ now represents a set of network events, e.g., TCP flags, and $\beta$ represents the set of network states. Finally, process discovery is applied to each $L_{1\dots\beta}\in\mathcal{L}_i$ to extract the set of $\beta$ process models $\mathcal{N}_{i}$. Each process model $N\in\mathcal{N}_{i}$ captures a distinct network state: the more network states, the finer-grained the pattern characterization is. Finally, pattern extraction also outputs the alignment distribution $\mathcal{D}_{i,1\dots\beta}$ for each state based on the diagnoses calculated between the event logs of $\mathcal{L}_i$ and $\mathcal{N}_i$ through alignment-based conformance checking (see Section \ref{sec:related_work}).

\subsection{Online Monitoring}
\label{OM}
With the BAG and process models in place, the system can be monitored and the risk of vulnerability exploitation evaluated dynamically. First, \textbf{traffic monitoring} captures traffic data from the different system nodes under unknown conditions. Such data are processed through \textbf{similarity evaluation}, which compares the set of process models for each $i$-th vulnerability $\mathcal{V}_i$. In particular, the traffic data associated with the $i$-th node is turned into a set of online event logs $\mathcal{L}_{O,i}$, which are checked against $\mathcal{V}_i$ with alignment-based conformance checking. The resulting online alignment distribution $\mathcal{D}_{o,i}$ is compared with the alignment distribution $\mathcal{D}_{i}$ obtained during offline system characterization through the cosine similarity:
\begin{equation*}
    \textrm{CosSim}(\mathcal{D}_{o,i}, \mathcal{D}_{i}) = \frac{\langle \mathcal{D}_{o,i}, \mathcal{D}_{i} \rangle}{||\mathcal{D}_{o,i}|| \, ||\mathcal{D}_{i}||}.
\end{equation*}
When $\textrm{CosSim}(\mathcal{D}_{o,i}, \mathcal{D}_{i})$ has higher values, a known exploitation is taking place, whereas low values of this metric indicate that the attack is not in course of action, i.e., the network traffic is legitimate. The resulting similarity is passed on to \textbf{BAG update}, which refreshes the CPT tables associated with the BAG using the similarity values of each network node. The updated BAG is leveraged by \textbf{dynamic risk assessment}, which executes Bayesian inference algorithms, such as Variable Elimination \cite{7885532}, thereby computing the posterior probability of an attacker reaching target nodes $P(v_{\text{target}}|v_{\text{attacker}})$, considering an adversary targeting the network as prior evidence.

%% file: sections/evaluation.tex
\section{Evaluation}
\label{sec:evaluation}
\begin{figure}[!t]
  \centering
  \includegraphics[width=\linewidth]{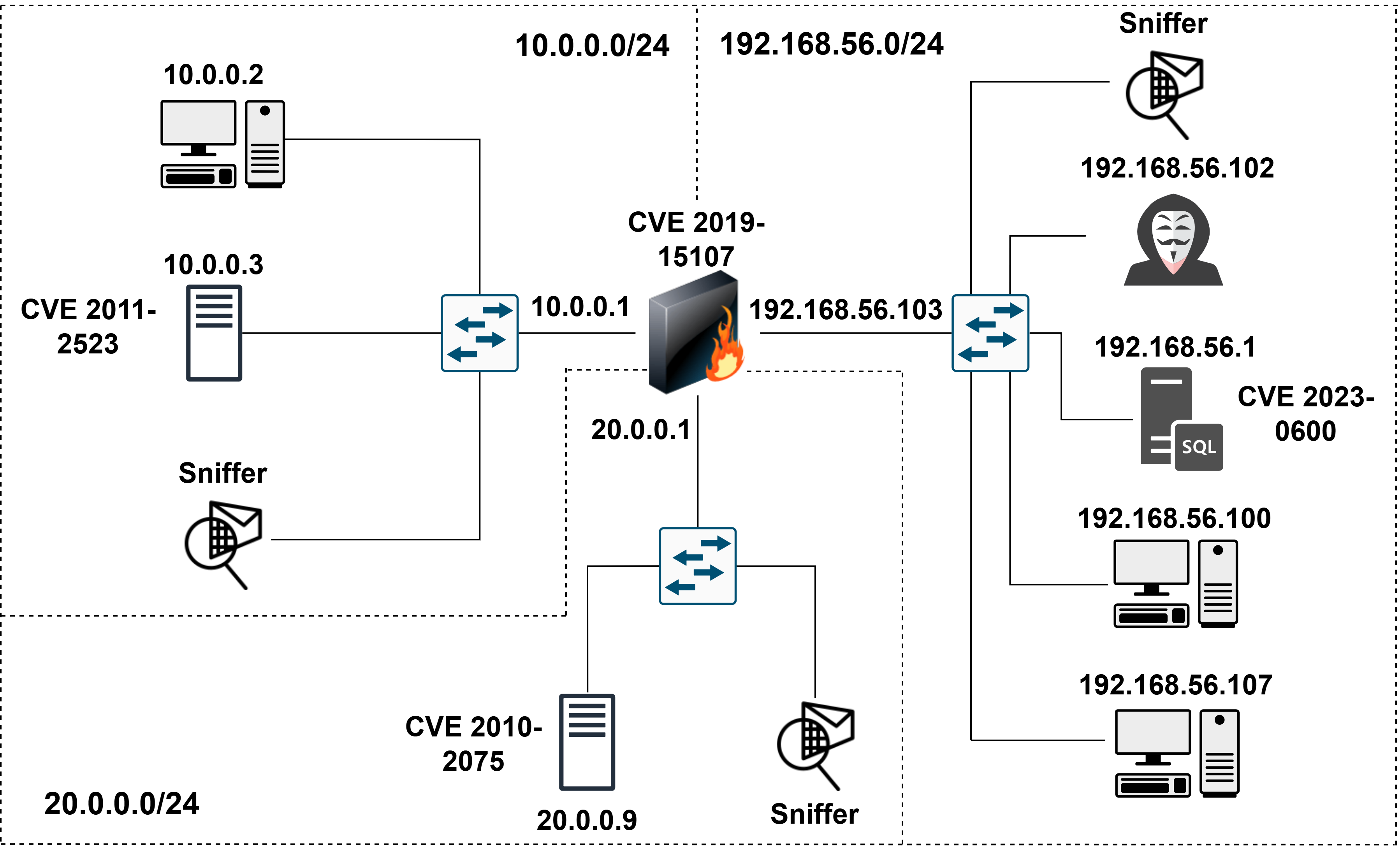}
  \caption{The cybersecurity testbed.}
  \label{fig:testbed}
\end{figure}
This section presents the application of our method to the cybersecurity testbed that we built for our experimentation, depicted in Fig. \ref{fig:testbed}, and shows the results of dynamic risk assessment with the instantiation of two attack paths.

\subsection{Offline System Characterization}
The adopted cybersecurity testbed consists of a corporate network infrastructure comprising three interconnected internal subnets linked via a Linux-based firewall. Specifically, subnet 192.168.56.0/24 represents the employee network, which is restricted from accessing the internal network (10.0.0.0/24) and can communicate solely with the DMZ (20.0.0.0/24); this subnet hosts employee-facing services such as web servers and VPN endpoints. The internal network (10.0.0.0/24) hosts critical servers containing sensitive assets, including patents, trade secrets, and proprietary data. Within each subnet, at least one device is affected by a known vulnerability, enabling an attacker who has gained access to the employee network to perform privilege escalation and lateral movement toward the high-value target at 10.0.0.3. In particular, the following well-known vulnerabilities were selected: (1) \texttt{CVE-2023-0600} affects host 192.168.56.1, enabling unauthenticated SQL injection in the ``WP Visitor Statistics (Real Time Traffic)" WordPress plugin (version $\leq 6.9$) for database credential extraction; (2) \texttt{CVE-2019-15107} impacts the firewall, allowing command injection in the Webmin management tool (version $\leq 1.920$) via \texttt{password\_change.cgi} for root Remote Code Execution (RCE); (3) \texttt{CVE-2010-2075} creates a backdoor in UnrealIRCd 3.2.8.1 on server 20.0.0.9, permitting unauthenticated remote command execution on TCP port 6667; (4) \texttt{CVE-2011-2523} (vsftpd 2.3.4 backdoor) on 10.0.0.3 opens a root shell on TCP port 6200 granting root access without authentication. Finally, in each subnet, a sniffer device is connected to capture real-time network traffic and store it in \texttt{pcap} files.

\begin{figure}[!t]
  \centering
  \includegraphics[width=0.75\columnwidth]{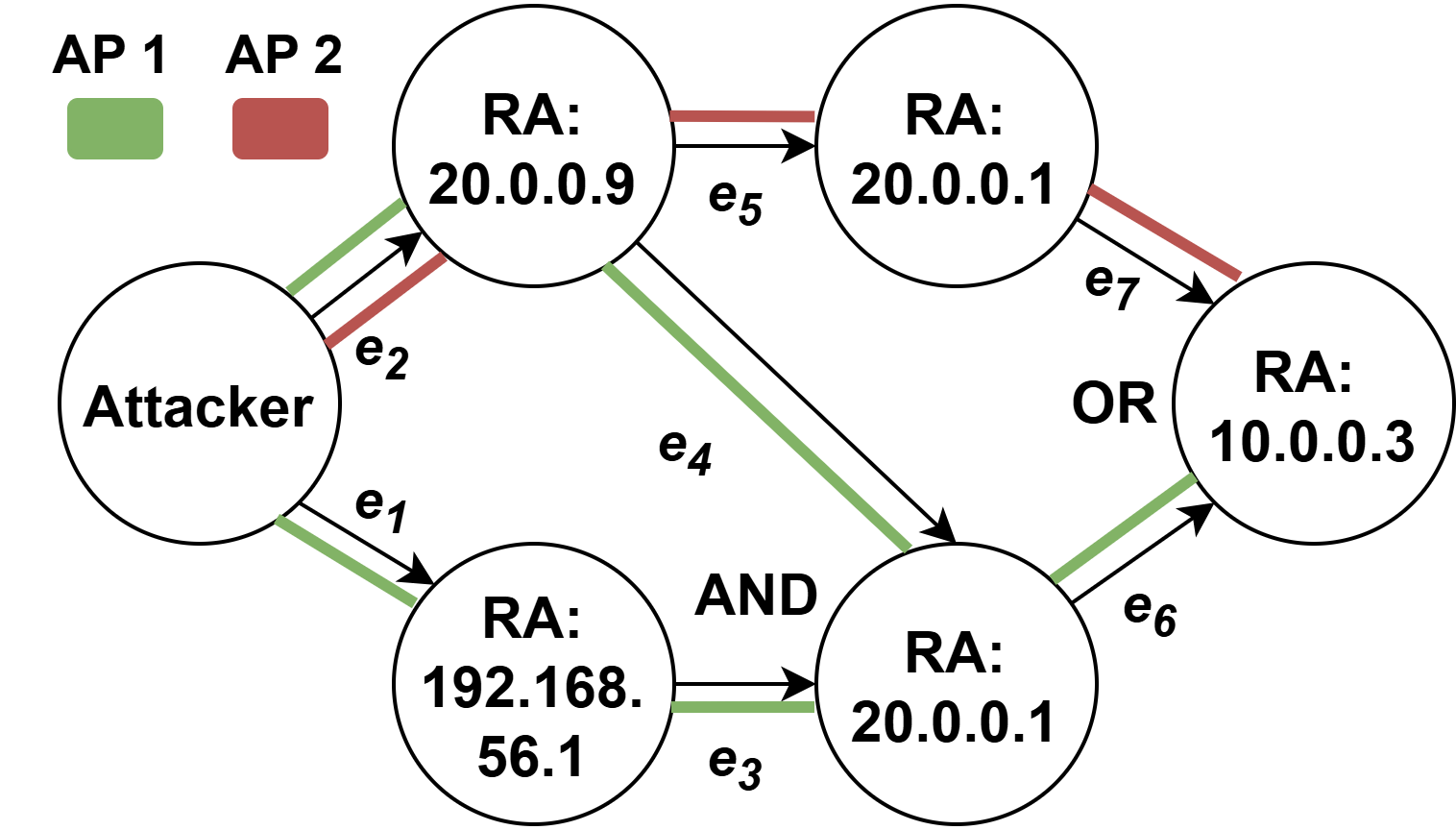}
  \caption{The BAG of the cybersecurity testbed. }
  \label{fig:BAG}
\end{figure}
\begin{table}[!t]
\fontsize{8pt}{9pt}\selectfont
\renewcommand{\arraystretch}{1.3}
\centering
\caption{Edge characterization in the BAG.}
\label{table:edges_mapping}
\begin{tabular}{c c c c}
\toprule
\bf Edge & $ v_i$ $\rightarrow$ $v_j$ & \bf Vulnerability $u$ \\
\midrule
$e_1$ & Attacker $\rightarrow$ RA:192.168.56.1 & CVE-2023-0600 \\
$e_2$ & Attacker $\rightarrow$ RA:20.0.0.9     & CVE-2010-2075 \\
$e_3$ & 192.168.56.1 $\rightarrow$ RA:20.0.0.1 (login) & Administrative Credentials \\
$e_4$ & RA:20.0.0.9 $\rightarrow$ RA:20.0.0.1 (login) & Administrative Credentials \\
$e_5$ & RA:20.0.0.9 $\rightarrow$ RA:20.0.0.1     & CVE-2019-15107 \\
$e_6$, $e_7$ & RA:20.0.0.1 $\rightarrow$ RA:10.0.0.3     & CVE-2011-2523 \\
\bottomrule
\end{tabular}
\end{table}

\begin{table}[ht]
    \centering
    \caption{Conditional Probability Table (CPT) for node $v_j$.}
    \label{tab:cpt}
    \begin{tabular}{lcc}
        \toprule
        \textbf{Node $v_i$} & \multicolumn{2}{c}{\textbf{Node $v_j$}} \\
        \cmidrule(lr){2-3}
         & \textit{False} & \textit{True} \\
        \midrule
        \textit{False}  & $1$ & $0$ \\
        \textit{True} & $1-\textrm{CosSim}(\mathcal{D}_{o,i}, \mathcal{D}_{i})$ & $\textrm{CosSim}(\mathcal{D}_{o,i}, \mathcal{D}_{i})$ \\
        \bottomrule
    \end{tabular}
\end{table}

Fig. \ref{fig:BAG} illustrates the BAG generated from the given network infrastructure and vulnerabilities affecting each device. In particular, the graph captures the logical sequence of steps an attacker, who gained access to the employee network (192.168.56.0/24), must execute to compromise the target device at 10.0.0.3. The characterization of each edge in terms of source and destination security condition, along with the associated enabling vulnerability, is detailed in Table \ref{table:edges_mapping}.
Moreover, Table~\ref{tab:cpt} reports the CPT associated with each node. As discussed in Section~\ref{OM}, the CPT is dynamically updated during the Online Monitoring phase based on the similarity value $\textrm{CosSim}(\mathcal{D}_{o,i}, \mathcal{D}_{i})$, which captures the likelihood of detecting a vulnerability exploitation in the network traffic. This probability is then encoded in the CPT, as it governs the transition from a source security condition node to a destination one.

According to the BAG, the attacker can pursue two distinct attack paths (AP1 and AP2) to reach the target. In both paths, the adversary must first acquire Root Access (RA) privileges on the DMZ server at 20.0.0.9, in order to compromise the firewall at 20.0.0.1, and remove its \texttt{iptables} rules to gain access to the internal network (10.0.0.0/24). Specifically, AP1 leverages vulnerabilities \texttt{CVE-2023-0600} and \texttt{CVE-2010-2075}: the former enables retrieval of firewall administrative credentials via SQL injection at the employee network
host (192.168.56.1), while the latter grants RA on the DMZ server through the UnrealIRCd service. By combining the credentials obtained from the employee machine with the network positioning provided by the DMZ server, the attacker can directly authenticate to the firewall's Webmin management interface, effectively bypassing the need for further exploitation. Once logged into the firewall, the attacker modifies the \texttt{iptables} rules to enable direct access to the internal network, ultimately exploiting \texttt{CVE-2011-2523} on 10.0.0.3 to secure root privileges. 
Alternatively, in path AP2, the attacker, after gaining RA privileges in the DMZ server, directly exploits \texttt{CVE-2019-15107} at the firewall in order to bypass the login credentials in the Webmin management interface and gain root privileges. Once the firewall is compromised, as in AP1, the attacker modifies the \texttt{iptables} rules to enable direct access to 10.0.0.3 and then exploits the local vulnerability (\texttt{CVE-2011-2523}) to attain root privileges on the target.

The exploitation of vulnerabilities for each AP has been executed through a Kali Linux machine connected to the employee network at IP address 192.168.56.102. Following the MITRE\&Attack framework, before executing the exploit, the attacker performs the following steps. First, the adversary conducts network scanning to map the target subnet and identify active hosts. Second, host scanning is performed on the discovered IP addresses to enumerate open ports and active services. Third, the attacker executes vulnerability scanning on these services to detect the presence of known security flaws (i.e., CVEs). Finally, upon confirming a vulnerable service, the attacker proceeds to the exploitation phase, deploying the appropriate payload to compromise the target and move laterally to the subsequent node in the attack graph.

Given $v_{1,\dots,4}$ the BAG nodes such that $v_1=\text{RA:192.168.56.1}$, $v_2=\text{RA:20.0.0.9}$, $v_3=\text{RA:20.0.0.1}$, and $v_4=\text{RA:10.0.0.3}$, our method extracts the sets of process models $\mathcal{V}_{1\dots,4}$ that characterize the anomalous traffic when the corresponding vulnerabilities are exploited during the instantiation of the two attack paths. To maximize the state separation within the $i$-th set of process models $\mathcal{V}_i$~\cite{vitale2025networktrafficanalysisprocess}, we split the network traffic into three states and obtain the alignment distribution $\mathcal{D}_{i,j}$ for each state $j$. Please, notice that BAG node $\text{RA:20.0.0.1 (login)}$ was not considered as it does not involve the exploitation of any vulnerability.

\subsection{Online Monitoring}

We have instantiated the two attack paths as follows. In both cases, we first stimulated the system nodes with legitimate traffic. Next, we progressively corrupted the network traffic node by node with the paths described above. This led to a multi-step dynamic risk assessment for each attack path. 

During traffic monitoring, the traffic data associated with nodes $v_{1,\dots,4}$ are collected with the packet sniffer. This produces the set of traffic data $\mathcal{T}_{1\dots4}$. Once the traffic data are collected, they are pre-processed to generate the four sets of online event logs $\mathcal{L}_{O,1\dots4}$, which are run through similarity evaluation to obtain the four similarity values $\textrm{CosSim}_{1\dots4}$. These values are used to update the CPTs of the BAG and obtain the final probabilities that the system nodes are compromised for dynamic risk assessment.

\begin{table}[!t]
\fontsize{8pt}{9pt}\selectfont
\renewcommand{\arraystretch}{0.75}
\centering
\caption{The CosSim values [$\%$] for each node per attack step and Attack Path (AP).}
\label{tab:results}
\begin{tabular}{cccccc}
\toprule
Attack & Node & \multicolumn{4}{c}{Attack Steps} \\
\cline{3-6}
& & I & II & III & IV \\ \midrule
\multirow{4}{*}{AP$_1$} & RA:192.168.56.1 & 10.0  & 99.9 & 99.9 & 99.9 \\
& RA:20.0.0.9 & 2.1  & 2.1  & 99.5 & 99.9 \\
& RA:20.0.0.1 & 0.3  & 0.3  & 0.3  & 2.1 \\
& RA:10.0.0.3 & 0.2  & 0.2  &      & 98.9 \\ \midrule
\multirow{4}{*}{AP$_2$} & RA:192.168.56.1 & 10.2 & 10.2 & 10.2 & 10.2 \\
& RA:20.0.0.9 & 2.1  & 99.5 & 99.9 & 99.9 \\
& RA:20.0.0.1 & 0.3  & 2.1  & 99.5 & 99.5 \\
& RA:10.0.0.3 & 0.2  & 0.2  & 0.2  & 98.9 \\ \bottomrule
\end{tabular}
\label{tab:cossim}
\end{table}

\subsection{Results}
Table \ref{tab:cossim} shows the \textrm{CosSim} values populating the CPT of each node in the BAG, updated for each attack step and attack path. For attack step I, the \textrm{CosSim} for each node is very low since only legitimate traffic circulates across nodes $v_{1\dots4}$. In the subsequent steps, malicious traffic that was replicated in the offline phase is progressively recognized. For example, for AP$_{1}$, the attack step II involves the compromise of node $v_1$ by leveraging vulnerability \texttt{CVE-2023-0600}; this is successfully recognized through the \textrm{CosSim} measure, which achieves 99.9\% similarity to the known malicious pattern. In the end, when attack paths are complete at attack step IV, the nodes involves in the attack show almost perfect similarity, outlining the utility of process mining in identifying malicious network traffic.

\begin{figure}[ht]
  \centering
  \includegraphics[width=\linewidth]{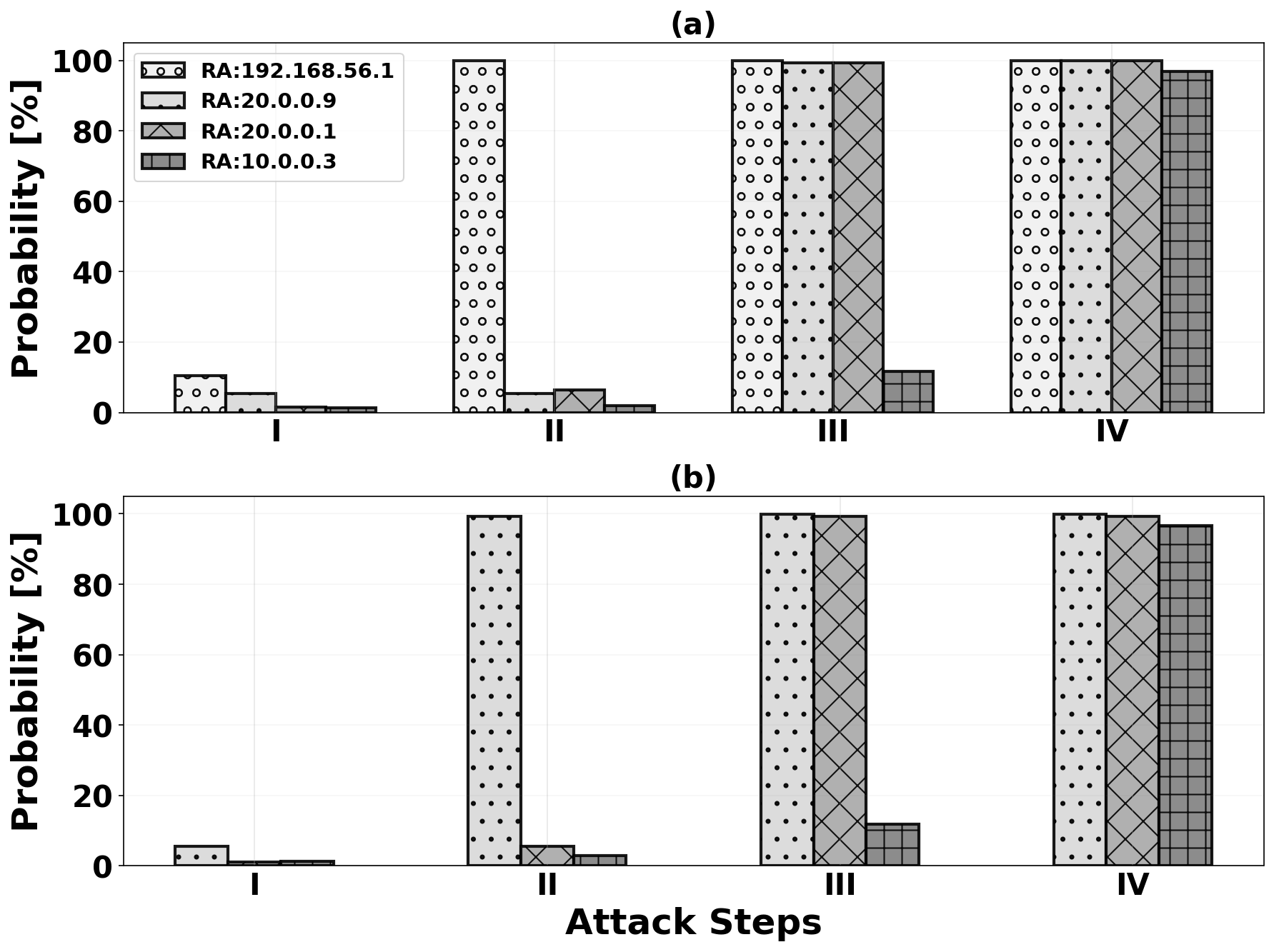}
  \caption{Exploit probability values for each attack step in AP1 (a) and AP2 (b).}
  \label{fig:res}
\end{figure}

Figure \ref{fig:res} shows the posterior compromise probabilities of all nodes along both attack paths, updated at each attack step. As the attacker traverses the attack path toward the target device, the posterior probability of compromise associated with each node increases, eventually reaching values close to 100\%. In particular, the compromise probability of the terminal node RA:10.0.0.3 rises at every step, reflecting the attacker’s progressive advancement toward the target device in the computer network. Specifically, in the first AP, the compromise probability increases from 1.2\% at Step I (when the attacker gains access to the employee network) to 1.8\% after the exploitation of \texttt{CVE-2023-0600} at Step II, and to 11.5\% after the exploitation of \texttt{CVE-2010-2075} and the subsequent login to the firewall at Step III. Finally, it reaches 96.8\% after the exploitation of \texttt{CVE-2011-2523}, which affects the target host. A similar trend is observed for the second attack path as well.

%% file: sections/conclusion.tex
\section{Conclusion}
\label{sec:conclusion}

This paper proposed a novel dynamic risk assessment method integrating BAGs and process mining, combining the insights of process-based packet-level inspection with the capabilities of BAGs. 
The proposed approach enables real-time updates to the probability of system compromise as evidence of vulnerability exploitation accumulates during an ongoing attack.
The method proves particularly valuable in two practical scenarios: honeypot deployments, where it supports attacker profiling and forensic analysis, and budget-constrained or legacy environments where patching is unfeasible, enabling timely response and mitigation actions based on exploitation probabilities along known attack paths.
We evaluated the approach on a cybersecurity testbed featuring a complex network topology and multiple vulnerable machines exposed to active cyber attacks.
Our results show that our method successfully characterizes malicious traffic monitored under the exploitation of vulnerabilities and allows dynamic risk assessment by updating a posteriori probabilities of the BAG with process mining-driven insights.

Nevertheless, the proposed method is subject to some limitations. First, exploitation probabilities at each node were estimated after a vulnerability was successfully exploited, without capturing the intermediate steps followed by the attacker during the exploitation process. Second, the approach was validated on a single case study, which may limit the generalizability of the results to larger networks. Third, no sensitivity analysis across different process mining algorithms was conducted, which leaves open the question of how algorithm choice affects the overall performance of the method.

Future work will include the extension of the proposal with intrusion detection systems able to automatically filter benign traffic and focus the analysis on malicious network data, which \revision{could enhance the ability of process mining to recognize the presence of anomalous packet-level network patterns associated with known vulnerabilities. In addition, we aim to} extend the proposal with automatic online mechanisms to account for past exploitation of system nodes.
